\documentstyle[epsfig,12pt,fullpage]{article}

\def\xf{x_{\scriptscriptstyle F}}
\def\nf{n_{\scriptscriptstyle F}}

\begin{document}
\begin{titlepage}
\begin{flushright}\vbox{\begin{tabular}{c}
           TIFR/TH/98-22\\
	   IMSc-98/06/33\\
           June, 1998\\
           hep-ph/9812409\\
\end{tabular}}\end{flushright}
\begin{center}
   {\large \bf
      Double Asymptotic Scaling in Drell-Yan Processes}
\end{center}
\bigskip
\begin{center}
   {Rahul Basu\footnote{E-mail: rahul@imsc.ernet.in}\\
    The Institute of Mathematical Sciences,\\
    Chennai (Madras) 600 113, India.\\
    \medskip and\\ \medskip
    Sourendu Gupta\footnote{E-mail: sgupta@theory.tifr.res.in}\\
    Department of Theoretical Physics,\\
    Tata Institute of Fundamental Research,\\
    Homi Bhabha Road, Mumbai 400005, India.}
\end{center}
\bigskip
\begin{abstract}
   Double scaling may be observed in the Drell-Yan process at the
   Tevatron and in the ALICE detector at the Large Hadron Collider
   in a window of masses. In the double scaling limit of the cross
   section, the higher order QCD corrections are quite large, and
   are driven by the rise in gluon densities in the double scaling
   regime. The naive parton model cannot be valid even at asymptotic
   energies in the region of phase space where double scaling holds.
\end{abstract}
\end{titlepage}

The HERA experiments have observed double asymptotic scaling \cite{ballforte} 
of the deep inelastic
structure function $F_2(x, Q^2)$ by probing a region of kinematics with large
momentum transfer, $Q^2$, and small Bjorken variable, $x$. In this
paper, we discuss the possibility of future hadron
colliders observing similar double scaling phenomena.
We examine the simplest such process here--- that of Drell-Yan
production of a lepton pair. The appropriate kinematic range will
be accessible to the ALICE experiment at the CERN Large Hadron
Collider and to experiments at the Tevatron.

In this process a lepton pair is produced in a hadron collider at a
centre of mass energy $\sqrt S$. The kinematics of the pair is specified
by its invariant mass, $Q$, and the longitudinal momentum fraction,
$\xf=2q_z/\sqrt S$. If $\xf=0$, then at large $\sqrt S$ and $Q$, 
the ratio
$\tau=Q^2/S$ can be small. The process will probe parton densities
in the double scaling region $x_1^0=x_2^0=\sqrt\tau\ll1$ and scale $Q^2$
large. The pair mass must satisfy the conditions $Q_0\le Q\le x_0\sqrt S$
with appropriate choices of $Q_0$ and $x_0$. In this region the higher
order QCD corrections are computable in the double asymptotic scaling
(DAS) \cite{ballforte} approximation
and are not negligible. This is what we show in this communication.

\begin{figure}[htb]
\vskip9truecm\centering
{\includegraphics{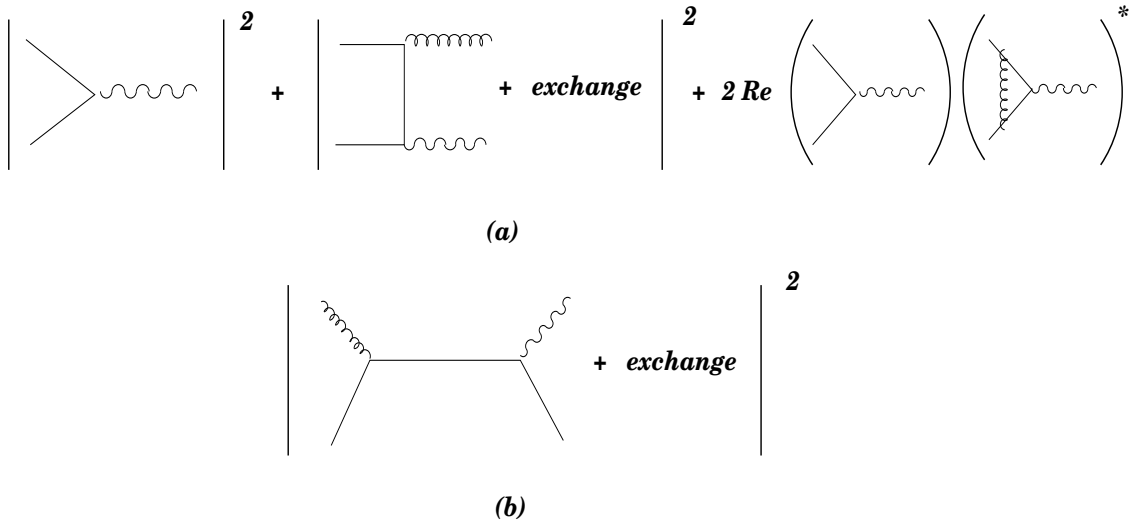}}
\caption{The diagrams that contribute to the Drell-Yan process up to
   $O(\alpha_s)$ in the strong coupling constant --- (a) the annihilation
   diagrams and (b) the Compton diagrams.}
\label{fg.diag}\end{figure}

The Drell-Yan cross section is the sum of the $\bar qq$ annihilation
process and the QCD Compton process \cite{dy}. The cross section at leading
order (LO) in QCD ({\em i.e} up to $O(\alpha_s)$ in the strong coupling
constant) comes from the diagrams shown in Figure \ref{fg.diag}. We treat
the two sets separately. Take the double differential cross section for
the annihilation contribution to the Drell-Yan cross section---
\begin{eqnarray}
  \nonumber
   {d\sigma_a\over d\log\tau d\xf}&=&
              {4\pi\alpha^2\over9S\sqrt{\xf^2+4\tau}}
       \biggl\{H(x_1^0,x_2^0,Q^2)
          \left[1+{2\alpha_s\over3\pi}\left({4\over3}\pi^2+1\right)\right]\\
  \nonumber
      & &\;\;
    +{2\alpha_s\over3\pi}\biggl[\int_{x_1^0}^1{dx_1\over x_1}
           H(x_1,x_2^0,Q^2)f_q\left({x_1^0\over x_1}\right)
    +\int_{x_2^0}^1{dx_2\over x_2}
           H(x_1^0,x_2,Q^2)f_q\left({x_2^0\over x_2}\right)\\
      & &\qquad\qquad
    +\int_{x_1^0}^1{dx_1\over x_1}
                                  \int_{x_2^0}^1{dx_2\over x_2}
           {H(x_1,x_2,Q^2)\over(1-z)(x_1+x_2)}
                     \tilde f_q(z,y^*) \biggr]\biggr\}.
\label{dyeq}\end{eqnarray}
The DAS approximation  must be consistently applied to the whole expression, 
{\sl i.e.\/},
to both the ``finite parts'',
\begin{eqnarray}
   f_q(z)&=&(1+z^2)\left({\log(1-z)\over1-z}\right)_+
      +{3\over2}\left({1\over1-z}\right)_+
      -2-3z,\\
   \tilde f_q(z,y^*)&=&-2(1-z)+{1+z^2\over(1-z)_+}
       \left[{1\over(1-y^*)_+}+{1\over y^*_+}\right],
\label{deff}\end{eqnarray}
and the parton density combination,
\begin{equation}
   H(x_1,x_2,Q^2)=\sum_fe_f^2\left[q_f(x_1,Q^2)\bar q_f(x_2,Q^2)
          +(x_1\leftrightarrow x_2)\right].
\label{defhg}\end{equation}

Before moving on to technicalities, we would like to set out the
justification for this claim. Recall that in defining the parton
densities through the DIS process, we absorbed collinear divergent 
parts of the higher order correction into the definition of these
densities through renormalisation group equations which are the
DGLAP equations. Of the collinear singular part, the terms proportional
to $\log(Q^2/\Lambda^2)$ give the splitting functions (Mellin transforms
of the anomalous dimensions of the leading twist operators) and the
remainder are called ``finite parts''. These latter are process
dependent. In the DAS approximation for deep-inelastic lepton scattering, 
the DGLAP
equations are solved analytically by expanding the anomalous dimensions
about their right-most singularity at $n=1$ in Mellin space {\sl i.e \/}
they are expanded in Mellin space in the quantity $\Delta=n-1$. The
leading terms are of the form $1/\Delta$ (and come entirely from the
splitting functions), and the first sub-leading terms are independent of
$\Delta$. In the Drell-Yan cross section in eq.\ (\ref{dyeq}) some of
the collinear divergent parts of the process have been absorbed into the
scale dependent parton density combination in eq.\ (\ref{defhg}). If this
is treated in the DAS limit, then certainly the remaining part of the collinear
divergent terms, in eq.\ (\ref{deff}), must be given the same treatment,
to wit the $1/\Delta$ and constant terms must be retained in the Mellin
transform (with respect to $\tau$) of eq.\ (\ref{dyeq}).

In order to be able to use the DAS forms of the quark and gluon
distribution functions in the double probability distributions given in
eq. \ (\ref{defhg}), we need to first express the quark distributions in
terms of the singlet (S) and non-singlet (NS) components. We define these
through the equations
\begin{equation}
   q_{\scriptscriptstyle S}=\sum_f^{\nf}(q_f +{\bar q_f})
      \qquad{\rm and}\qquad
   q_{{\scriptscriptstyle NS},f}= 2\nf q_f - q_{\scriptscriptstyle S}.
\end{equation}
Here $\nf$ is the number of active flavours. This allows us to express the
quark density for flavour $f$, $q_f$, in terms of $q_{\scriptscriptstyle S}$
and $q_{{\scriptscriptstyle NS},f}$. In the DAS limit where the non-singlet
can be neglected (up to subleading order) compared to the singlet--- $q_f$ is
just given by the singlet quark densities.

Thus in the DAS limit the DGLAP equations can be solved exactly to give
\begin{eqnarray}
{\cal G}(x,Q^2)&=&xg(x,Q^2)
	 =\left({N^2\over4\pi\gamma\sigma}\right)^{1/2}
      \exp\left[2\gamma\sigma-\delta_+{\sigma\over\rho}\right]
         \left\{1+{\cal O}(1/\sigma)\right\},\\
Q(x,Q^2)&=& xq_S(x,Q^2) = N^\prime {\cal G}(x,Q^2) \gamma/\rho.
\label{ballf}\end{eqnarray}
where $\rho\sigma=\log(x_0/x)$ and $\sigma/\rho=\log(t/t_0)$ with $t=\ln
(Q^2/\Lambda^2)$ ($t_0$ which is $t$ evaluated at $Q_0^2$,  and
$x_0$ are chosen so that the approximation holds). The usual gluon density
function is denoted by $g(x,Q^2)$. The constants appearing
in these equations are given by
\begin{equation}
   \gamma^2={12\over\beta_0},\quad
   \delta={1\over\beta_0}\left(11+{2\over27}\nf\right),\quad
   \beta_0=11-{2\over3}\nf,
\label{cballf}
\end{equation}
and the running coupling $\alpha_s(t)=4\pi/(\beta_0 t)$.
$N$ and $N^\prime$ are overall
normalisation constants which will be fitted with data or
parametrisations. These are expressions valid to lowest order in the
strong coupling constant $g_s$.

There are no $1/\Delta$ terms in the Mellin transform of $f_q$ and $\tilde
f_q$. The constant terms are identifiable by taking the first moment of
these quantities---
\begin{equation}
   M[f_q]=-{7\over2}+{\cal O}(\Delta),\qquad{\rm and}\qquad
   M[\tilde f_q]=-1+{\cal O}(\Delta).
\label{mellinf}\end{equation}
As a result, the leading part of the finite terms can be modelled by
the $z$-space functional forms---
\begin{equation}
   f_q(z)\simeq-{7\over2}\delta(1-z)\qquad{\rm and}\qquad
   \tilde f_q(z,y^*)\simeq-\delta(1-z).
\label{subleadinf}\end{equation}
The delta-functions tell us that this subleading correction is due
entirely to the soft part of the phase space in the QCD corrections.

Substituting these into eq.\ (\ref{dyeq}) then gives us---
\begin{equation}
   \left.{d\sigma_a\over d\log\tau d\xf}\right|_{\xf=0}=
          {2\pi\alpha^2H(\sqrt\tau,\sqrt\tau,Q^2)\over9S\sqrt\tau}
       \left[1+{2\alpha_s\over3\pi}
          \left({4\over3}\pi^2-6-{1\over2\sqrt\tau}\right)\right].
\label{final}\end{equation}
The appropriate expression for the parton density combination is
\begin{equation}
   H\left(x_1^0,x_2^0,Q^2\right)= H_0
         \left({1\over4\pi\sqrt{\sigma_1\sigma_2}}\right)
      \exp\left[2\gamma(\sigma_1+\sigma_2)
          -2\delta\log(t/t_0)\right]\frac{\gamma}{\rho^2 x_1^0x_2^0},
\label{bfh}\end{equation}
where $x_1^0=x_2^0=\sqrt\tau$, and hence $\sigma_1=\sigma_2=\log(x_0/
\sqrt\tau)$. The overall normalisation constant $H_0$ can be fixed from
data on parton densities.

We turn next to the contribution of the Compton process to the cross
section. The full expression is---
\begin{eqnarray}
  \nonumber
   {d\sigma_c\over d\log\tau d\xf}&=&
              {\alpha^2\alpha_s\over9S\sqrt{\xf^2+4\tau}}
    \biggl\{\biggl[\int_{x_1^0}^1{dx_1\over x_1}
           G(x_1,x_2^0,Q^2)f_g\left({x_1^0\over x_1}\right)
    +\int_{x_2^0}^1{dx_2\over x_2}
           G(x_2^0,x_1,Q^2)f_g\left({x_2^0\over x_2}\right)\\
  \nonumber
      & &\;\;
    +\int_{x_1^0}^1{dx_1\over x_1}\int_{x_2^0}^1{dx_2\over x_2}
           {G(x_1,x_2,Q^2)\over(1-z)(x_1+x_2)}
                     \tilde f_g(z,1-y^*)\\
      & &\qquad\qquad
    +\int_{x_1^0}^1{dx_1\over x_1}\int_{x_2^0}^1{dx_2\over x_2}
           {G(x_2,x_1,Q^2)\over(1-z)(x_1+x_2)}
                     \tilde f_g(z,y^*)\biggr\}
\label{dyeqg}\end{eqnarray}
where the finite parts,
\begin{eqnarray}
   f_g(z)&=&[z^2+(1-z)^2]\log(1-z)+1-6z(1-z),\\
   \tilde f_g(z,y^*)&=&2z(1-z)+(1-z)^2y^*+[z^2+(1-z)^2]{1\over y^*_+},
\label{fing}\end{eqnarray}
and the parton density combination,
\begin{equation}
   G(x_1,x_2,Q^2)=g(x_2,Q^2)\sum_f e_f^2\left[
        q_f(x_1,Q^2)+\bar q_f(x_1,Q^2)\right]
\label{pang}\end{equation}
are both to be evaluated consistently in the DAS limit.

The leading parts of the finite terms are identifiable from the first
moment---
\begin{equation}
   M[f_g]=-{13\over18}+{\cal O}(\Delta),\qquad{\rm and}\qquad
   M[\tilde f_g]={1\over3}\left(1 + y^*+{2\over y^*_+}\right)+{\cal O}(\Delta).
\label{melling}\end{equation}
In $z$-space, they can be modelled up to subleading order in $\Delta$ by the
functions
\begin{equation}
   f_g(z)=-{13\over18}\delta(1-z)\qquad{\rm and}\qquad
   \tilde f_g(z,y^*)=\left[{1\over3}\left(1 + y^*+{2\over y^*_+}\right)
           \right]\delta(1-z).
\label{subleading}
\end{equation}
Substituting into eq.\ (\ref{dyeqg}) gives the DAS result---
\begin{equation}
   \left.{d\sigma_c\over d\log\tau d\xf}\right|_{\xf=0}=
          {2\pi\alpha^2\alpha_sG(\sqrt\tau,\sqrt\tau,Q^2)\over9S\sqrt\tau}
      \left[-{13\over9}+{1\over2\sqrt\tau}\right].
\label{ginal}\end{equation}
The appropriate expression for the parton density combination is
\begin{equation}
 G\left(\sqrt\tau,\sqrt\tau,Q^2\right)=G_0
         \left({1\over4\pi\gamma\sigma}\right)
         \exp\left[-2\delta\log{t\over t_0}\right]
      \left[{\gamma\over\rho\sqrt\tau}\exp[4\gamma\sigma]
      \right].
\label{bfg}\end{equation}
The constant $G_0$ can again be fixed from the numerical values of the
parton densities.

This concludes the computation, since the Drell-Yan cross section
\begin{equation}
   \left.S{d\sigma_a\over d\log\tau d\xf}\right|_{\xf=0}\;=\;
   \left.S{d\sigma_a\over d\log\tau d\xf}\right|_{\xf=0}
    +\left.S{d\sigma_c\over d\log\tau d\xf}\right|_{\xf=0}
\label{total}\end{equation}
and the two terms on the right are given in
eqs.\ (\ref{final},\ref{ginal}). We make three
remarks---
\begin{itemize}
\item
   Since $\xf=0$, we have $x_1^0=x_2^0=\sqrt\tau$, and hence 
   \begin{equation}
      \sigma_1=\sigma_2= \sqrt{\log(t/t_0)\log(x_0/\sqrt\tau)}.
   \end{equation}
   Also, $t=\log(S\tau/\Lambda^2)$. Hence, the only free variable in
   eqs.\ (\ref{final},\ref{ginal}) is $\tau$.
\item
   In general, QCD corrections involve parton densities at
   all momentum fractions in the range $\sqrt\tau<x<1$, whereas in the
   DAS only the densities at $x=\sqrt\tau$ are involved. This is
   due to the delta-functions in eqs.\ (\ref{subleadinf},\ref{subleading}).
   The same physics reduces the DGLAP equations for structure functions to
   the wave equation \cite{ballforte}.
\item
   The coefficient of $\alpha_s$ in eq.\ (\ref{final}) is negative and
   larger than unity. This is not a problem, since this
   cross section is not physical, being only one part of the total
   physical cross section. Adding eq.\ (\ref{ginal}) makes the total
   physical cross section positive.
\end{itemize}

Since we examine DAS at leading loop order, the parton densities to be
used in fixing the normalisations $H_0$ and $G_0$ must be obtained at
LO. For this reason we work with the GRV 94 LO \cite{grv} set of parton
density parametrisations. We find that DAS is valid if we choose
$Q_0^2=1{\rm\ GeV}^2$ and $x_0=0.1$. It turns out that $G_0/H_0\approx40$.
Such a large value of this ratio is generic, since the gluon densities are
substantially larger than the quark densities. As a result, the Compton
process dominates over the annihilation process in the DAS.

Recall that the increase of the Drell-Yan K-factor over unity is due to
the parts of the cross section proportional to $\alpha_s$. From the DAS
expressions, and the fact that $G_0\gg H_0$, it is clear that the K-factor
is much larger than unity for $\sqrt S\to\infty$. In fact, from the behaviour
of $\alpha_s$, one can see that the K-factor vanishes as $1/t$. The
rest of the factors in eqs.\ (\ref{final},\ref{ginal}) are almost
independent of $S$ at fixed $\tau$. Since the K-factor in the DAS remains
non-negligible at the largest accessible energies, the naive parton
model cannot be recovered wherever DAS is valid. In contrast, the K-factor
at larger values of $\tau$ are rather close to unity at these energies
\cite{hard}, and the parton model may be taken as a reasonable approximation
to reality.

\begin{figure}[htb]
\vskip9truecm\centering
{\includegraphics{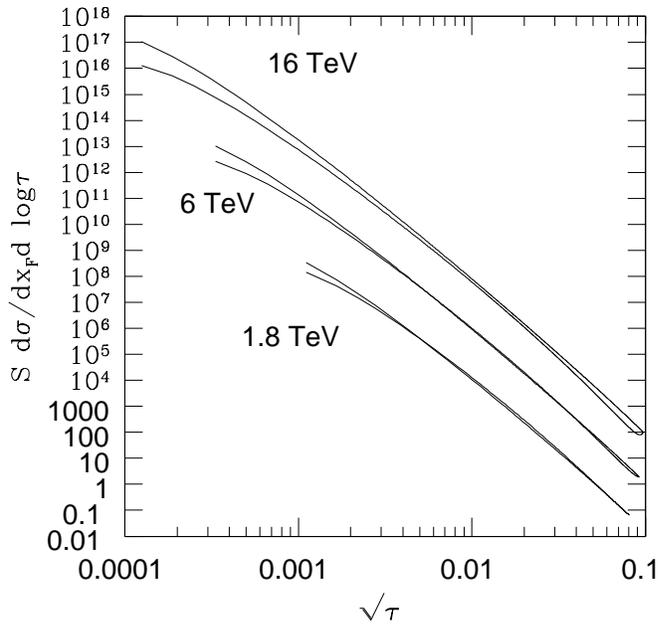}}
\caption{The DAS prediction for the Drell-Yan cross section as a
    function of $\protect{\sqrt\tau}$ for $\xf=0$ compared with the 
    full cross section at three different values of $\protect\sqrt S$.
    In each case, the DAS curve is the lower one. The curves for 16 TeV
    and 1.8 TeV have been scaled by a factor of 100 and .01 respectively
    for clarity in display.}
\label{fg.comp}\end{figure}

In Figure \ref{fg.comp} we compare the DAS forms of the cross section with
the full one-loop expression. It is interesting to note the disagreement
between the two sets of computations at low $\tau$. This is expected,
since low $\tau$ corresponds to low $Q^2$, where DAS breaks down.
Similarly, at large $\tau$, DAS breaks down because $x$ is not small
enough. Thus, the applicability of DAS to Drell-Yan is bounded both above
and below, in $\tau$. This is to be contrasted with the situation in
DIS.

\begin{figure}[htb]
\vskip 9truecm\centering
{\includegraphics{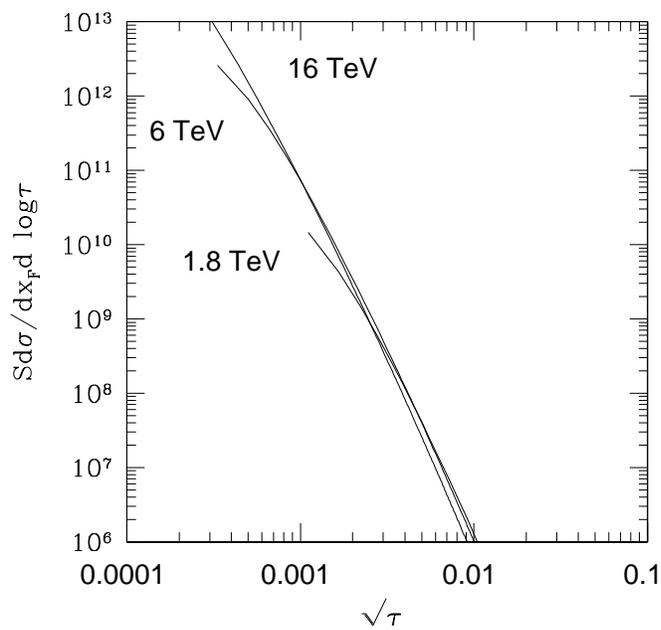}}
\caption{The DAS prediction for the Drell-Yan cross section as a
    function of $\protect{\sqrt\tau}$ for $\xf=0$ at the values of 
    $\protect{\sqrt S}$ shown.}
\label{fg.das}\end{figure}

In Figure \ref{fg.das} we show the differential cross section $Sd\sigma/
d\log\tau dx_F$ as a function of $\sqrt\tau$ for three different values
of $\sqrt S$.
Note that the cross section for these three values of $\sqrt S$ lie
almost on one single universal curve. This is the double scaling
curve. Departures from DAS are visible at $\sqrt\tau\simeq x_0$ and
$\tau S\simeq Q_0^2$.

In the DAS limit of hadronic cross sections, the leading correction is
universal--- this is the $1/\Delta$ piece which comes entirely from the
splitting functions, {\sl i.\ e.\/}, the collinear region of phase space,
and is absorbed into the parton densities. The next term in the expansion
in $\Delta$ is not universal. However, since it arises in the soft part
of phase space, it should be easier to evaluate and sum the DAS phase
space to all loop orders than to evaluate the full correction. In fact,
this resummation might even be necessary. The reason is in the large
value of the K-factor, as mentioned above. Since the one loop correction
is so large, it may actually be necessary to sum the $\log Q^2\log(1/\tau)$
contributions to all orders, so as to describe the physical process. This
is an extension to hadronic collision of the usual DLL procedure in
lepton-proton DIS. We leave this for the future.

One of us (RB) would like to thank D. Indumathi for useful discussions.
This work was started at the 5th Workshop on High Energy 
Physics Phenomenology (WHEPP-5), held in January 1998 at Pune, India.

\end{document}